\documentclass[aps,preprint,eqsecnum,nofootinbib]{revtex4}

\usepackage{epsfig}

\newcommand{\be}{\begin{equation}}
\newcommand{\ee}{\end{equation}}
\newcommand{\bea}{\begin{eqnarray}}
\newcommand{\beas}{\begin{eqnarray*}}
\newcommand{\eea}{\end{eqnarray}}
\newcommand{\eeas}{\end{eqnarray*}}
\newcommand{\ba}{\begin{array}}
\newcommand{\ea}{\end{array}}
\def\ls{\mathrel{\lower4pt\vbox{\lineskip=0pt\baselineskip=0pt
           \hbox{$<$}\hbox{$\sim$}}}}
\def\gs{\mathrel{\lower4pt\vbox{\lineskip=0pt\baselineskip=0pt
           \hbox{$>$}\hbox{$\sim$}}}}

\begin{document}

\title{Radion stabilization from the vacuum on flat extra dimensions.}

\author{El\'{\i} Santos~$^{a,b}$ \footnote{email: eli@unach.mx},
 A. P\'erez-Lorenzana~$^{c}$ \footnote{email: aplorenz@fis.cinvestav.mx}
 and Luis O. Pimentel~$^{a}$ \footnote{email: lopr@xanum.uam.mx}}

\affiliation{
$^{a}$~Departamento de F\'isica, Universidad Aut\'onoma Metropolitana. \\
Apdo. Post. 55-534, C. P. 09340 M\'exico, D.F., M\'exico\\
$^{b}$ Secretar\'{\i}a Acad\'emica de F\'isica y  Matem\'aticas, 
Fac. de Ingenier\'ia, Universidad Aut\'onoma de Chiapas\\
Calle $4^a$ Oriente Nte. 1428, Tuxtla Guti\'errez, Chiapas, M\'exico \\ 
$^{c}$~Departamento de F\'{\i}sica, Centro de Investigaci\'on y de Estudios
Avanzados del I.P.N.\\
Apdo. Post. 14-740, 07000, M\'exico, D.F., M\'exico}

%\date{May 2007}

\begin{abstract}
Volume stabilization in models with flat extra dimension could follow from
vacuum energy residing in the bulk  when translational invariance is
spontaneously broken. We study a simple toy model that exemplifies this
mechanism which considers a massive scalar field with non trivial boundary
conditions at the end  points of the compact space, and includes contributions
from brane and bulk cosmological constants. We perform our analysis   in the
conformal frame where the radion field, associated with volume variations,  is
defined, and present a general strategy for building stabilization potentials
out of those ingredients. We also provide working examples for the interval and
the  $T^n/Z_2$ orbifold configuration. 
\end{abstract}
%\keywords{Field Theories in Higher Dimensions; Large Extra Dimensions}
%\packs{11.10.Kk }
%\vskip90pt
\maketitle
%%%%%%%%%%%%%%%%%%%%%%%%%%%%%%%%%%%%%%%%%%%%%%%%%%%%%%%%%%%%%%%%%%%%%%
\section{Introduction}

Extra compact spatial dimensions are a well known fundamental
ingredient  of String Theory, which needs  to be formulated at least
in ten dimensions (or eleven for M-theory) to be  consistent. The
space generated by the six (seven) space-like extra dimensions may
have a non trivial configuration and topology, and be characterized by a
variety of sizes, that, according to some speculations~\cite{dvali},
may even be as large as few micrometers, in contrast with the much
smaller Planck length, $\ell_P\sim ~ 10^{-33}$~cm. The idea seems to
find some motivation from the study of  the non-perturbative regime of
the $E_8\times E_8$ theory by Witten and Horava~\cite{witten}, where
one  of these extra dimensions appears to  be larger than the naively
expected Planck size for quantum gravity physics. 
The possibility that
there could be such extra dimension has renewed the interest in a
class of models once inspired by the works of Kaluza and
Klein~\cite{kk}; and lately suggested by several authors~\cite{rubakov,anto}.
It has also motivated a  large number of studies oriented to explore
phenomenological uses  of such new dimensions.

Of particular interest are the so called brane models, in which  our
observable world is constrained to live on a four dimensional
hypersurface (the brane) embedded in a flat higher dimensional space
(the bulk), such that the extra dimensions can only be tested by
gravity, and perhaps standard model singlets, a set up that resembles
D-brane theory constructions. Further modifications to this basic
scenario have also considered the possibility that some or even all
standard model fields may probe some of the extra dimensions. Nonetheless,
most models are studied only in the effective field theory limit,
valid below the fundamental string scale. These models have the extra
feature that they may provide an understanding of the large difference
among Planck, $M_P$, and electroweak, $m_{ew}$, scales almost by
construction, since now Planck scale ceases to be fundamental. It is
replaced by the truly fundamental gravity scale, $M_\ast$, associated
to quantum gravity in the $(4+n)$ dimensional theory. Both scales are
then related by the volume of the compact manifold, $vol_n$, through
out the expression~\cite{dvali}
 \be
 M_{P}^2 = M_\ast^{n+2}\,vol_n~,
  \label{add0}
 \ee
which indicates that the so far unknown value for $M_\ast$ could lay
anywhere within $m_{ew}$ and $M_P$. If it happens to be in the TeV
range there would be no big hierarchy, but a rather large volume is
required. A number of possible theoretical uses of such extra
dimensions had been explored, including new possible ways for the
understanding of mass hierarchies~\cite{xdhierarchy}, the origin of
neutrino masses~\cite{xdnumodels}, the number of matter
generations~\cite{dobrescu}, baryon number
violation~\cite{xdbviolation}, the origin of dark matter~\cite{xddm},
new mechanisms for symmetry breaking~\cite{xdsm} and model
building~\cite{xdmodels}, among many others. Experimental implications
of some of those models have been also under the scope of many
investigations (for references, see for instance~\cite{expt,coll}).

Most phenomenological models built on this scenario usually assume
that the extra dimensions are stable, which typically becomes  a
fundamental requirement since most effects of  extra dimensions on
low energy physics depend either on the effective size of the compact
space $b_0\sim vol_n^{1/n}$ or the effective Planck scale. However, if
the compact space were dynamical those quantities would become time dependent,
against observations.

As it can be easily realized, during the early inflation period and
the later evolution of the Universe, a static bulk appears to be hard to 
hard to accept against an expanding 4D word. Indeed, there are indications
that the contrary is rather more likely to happen~\cite{mmp}. 
Inflaton energy may also induce dynamical effects  on the extra space, by
driving the so called radion field,  associated with the overall extra volume
variations,  beyond its desired stable point. This may particularly
affect the large extra dimensional case where inflaton
contribution might increase the effective volume by a factor of few.

Understanding the stability of the compact space can be seeing as
finding the mechanism that provides the force which keeps the radion
fixed at its zero value. Thus, in order to have a stable bulk volume,
there has to be a potential which provides such a force. Some ideas on
the possible origin of this potential can be found in the literature,
ranging from pure quantum effects to String theory non trivial flux
constructions, see for
instance~\cite{kaluza-klein,tsujikawa,kklt,Frey,Joe1,quiros,wise,maru,chacko}. 
In this paper
we explore an idea first introduced in the context of warped extra
dimensions~\cite{wise}, and latter discussed for a single flat extra
dimension in Ref.~\cite{chacko}, where  vacuum energy is regarded as
the one responsible for generating the stabilization potential. 
Although the use of this mechanism on flat backgrounds
might look trivial at first sight, 
we believe an extended and careful analysis
is worthy for two reasons. First of all,
the mechanism on flat compact space mimics the
stringy scenario where fluxes are used for stabilization, 
providing a bottom-up toy model where other problems,
as metric back reactions and quantum stability could be tested.
Secondly, as we shall observe, the definition of the radion field 
on the frame where gravity action becomes standard implies the 
introduction of conformal factors on matter actions, which 
have a non trivial impact on  the stabilization analysis. 
This is a feature that has been usually overlooked in previous works (see for
instance~\cite{chacko}).

We will
consider the generic model where a massive scalar bulk field develops
a vacuum configuration that explicitly violates translational
invariance along the extra space. Such a vacuum, in the effective four
dimensions,  appears as a potential energy that depends on the size
of the extra dimension,  and thus it is interpreted as a radion potential.  
We argue that these potentials can be build to have 
a minimum at a finite and nonzero value of the extra dimensional size,
providing a successful stabilization  at  the tree level of
the theory.
Our analysis will concentrate mainly on the  
phenomenological modeling for the stabilization potential  on  flat
backgrounds, 
which,  given the number of particle
physics models built  on such an
assumption~\cite{xdhierarchy,xdnumodels,dobrescu,xdbviolation,xddm,xdsm,xdmodels},
we believe has some interest on their own.
Such an  approximation, 
however, would lack  the immediate link with the more
fundamental string theory that motivate it. And although the ingredients we
shall considerer are the minimum we expect  to come from a real string
theory construction, this is an issue  we will not address in here. 
Our main
goal will be to demonstrate in a practical constructive way that vacuum energy
could be enough to provide the required bulk stabilization, and to keep things
simple, we will work in the assumption that backreactions are negligible. 

The paper is organized as follows. First, we  discuss the general
aspects of radion stabilization by  vacuum energy on flat extra
dimensions. To clarify the basis of the mechanism, 
we start by briefly reviewing the definition of the radion
field, conformally mapping the initial action to the physical Einstein frame,
where 4D gravity action is kept as usual, and gravity coupling remains constant. 
We discuss the effect of such a metric conformal transformation  on other
Lagrangian terms on the action, particularly, on those that would later
contribute to  the stabilization potential. 
We show that, in general, when
the radius is away from its stable value, some conformal factors 
remain on the potential energy. Such factors
define the couplings of the radion to matter fields,
both in the bulk and on the brane. They also affect the stabilization
potential by introducing overall inverse volume factors.  
In section three, we discuss the mechanism for radion stabilization based on
vacuum energy. We first show that, 
surprisingly, 
stabilization can be accomplished with the sole introduction of cosmological
constants, which exemplifies the non trivial features of the mechanism.
Next, we shall consider brane and bulk cosmological constants as well
as bulk scalar vacuum contributions. We provide some general guidelines for
building a successful radion potential, which aside of 
having a non trivial minimum, may also insures a zero
4D effective cosmological constant in the Einstein frame.
Finally, in section four, we investigate the
implementation of the present mechanism  for the interval and for
$T^n/Z_2$ orbifolds.
For these examples, we show that brane and bulk cosmological constants play an
important role to control the profile of the stabilization potentials.  We end
with some concluding remarks and observations.

%%%%%%%%%%%%%%%%%%%%%%%%%%%%%%%%%%%%%%%%%%%%%%%%%%%%%%%%%%%%%%%%%%%%%
\section{The radion in the Einstein frame}

\subsection{Dimensional reduction and the radion field}

We start the discussion by
assuming that Einstein gravity holds in the
complete $(4+n)$D theory, and proceed with
dimensional reduction to introduce the definition
of the radion field and its couplings.
Thus we first write down
the Einstein-Hilbert action
\be
 S=
 \frac{ M_{\ast}^{2+n}}{2}\, \int\!d^4x\, d^n y\,
 \sqrt{|g_{(4+n)}|}\, R_{(4+n)}
 \ee
where  $R_{(4+n)}$
stands for the $(4+n)$D dimensional  scalar curvature, and
$|g_{(4+n)}|$ is the absolute determinant of the $(4+n)$D metric.
We then consider the background metric parameterization
$ ds^2 = g_{AB}dx^A dx^B =  g_{\mu\nu}dx^\mu dx^\nu - h_{ab}dy^a dy^b$,
that is conformally
consistent with 4D Poincar\`e invariance, and describes a compact and flat extra
space. So we assume
$y^a$ as dimensionless coordinates on a unitary and closed manifold.
Thus, $h_{ab}$
has length dimension two. Here we use  for the indices the convention
$A,B = \mu,a$ where $\mu = 0,\dots,3$ and $a=5,\dots, 4+n$.
Notice  we are not considering the usual  vectorlike $A_\mu^a$ connection
 pieces. This is so because we want to concentrate
only on the variations  of the metric along the transverse directions
for the rest of our discussion.

Upon dimensional reduction,  one  obtains at the zero mode level
 \be
 S=
 \frac{M_P^2}{2}\, \int\!d^4x\,
 \sqrt{|g_{(4)}|}\, \frac{\sqrt{|h|}}{vol_n}\,
 \left \{ R_{(4)} - \frac{1}{4} \partial_\mu h^{ab}\, \partial^\mu h_{ab}
 -\frac{1}{4}h^{ab}\partial_\mu h_{ab}\cdot h^{cd}\partial^\mu h_{cd}
 \right \}\,,
 \label{rg1}
 \ee
where
$vol_n$  stands for the volume of the extra space
at the desired stable configuration, as defined above in Eq.~(\ref{add0}).
In this initial frame gravity is not well defined. There is an 
extra factor which is in general different from unity when the compact volume
differs from that of $vol_n$. In order to get a proper gravity action one has to
go to a different frame.  Thus, we perform the conformal transformation
 \be
 g_{\mu\nu}\rightarrow e^{2\varphi}~g_{\mu\nu}\,,
 \label{conformal}
 \ee
with $e^{2\varphi}={vol_n}/{\sqrt{|h|}}$, to obtain
the $4$D gravity in canonical form
 \be
 S=
 \frac{M_P^2}{2}\, \int\!d^4x\,\sqrt{|g_{(4)}|}\,
 \left \{  R_{(4)} - \frac{1}{4} \partial_\mu h^{ab}\, \partial^\mu h_{ab}
 + \frac{1}{8}h^{ab}\partial_\mu h_{ab}\cdot h^{cd}\partial^\mu h_{cd}
 \right \}\,,
 \label{rg2}
 \ee
in what we shall refer as the conformal (or Einstein) frame. Next,
$g_{\mu\nu}$ can be assumed to be the standard  metric  for a Poincar\'e
invariant brane Universe or the Friedmann-Robertson-Walker metric for
cosmology. We will, however, keep $g_{\mu\nu}$ undefined  as far as
possible. Nevertheless, to simplify, we shall take $ h_{ab} =
b^2\delta_{ab}$, such that $b$
represents the actual size of the compact space. 

If the bulk had the
desired stable configuration, the physical size of
the extra dimension would be given by the identification $b= b_0$, such that 
$vol_n=b_0^n$.
However, in cosmological grounds at least, 
it is plausible that  $b$ would be a
time dependent field, thus, the actual physical volume of the bulk
would rather be  given as $vol_{\rm phys}=\sqrt{|h|}= b^n(t)$. A more
general dependence $b(x)$ on the four space time coordinates may also
be possible.  This would describe local variations on the bulk radius
along the brane. Although we will not explicitly refer to this case in
here, we will kept most  expressions as general as possible.

As it can be read from the action, 
in the conformal frame the effective Planck scale is well defined and
constant. However, volume variation effects appear as the scalar field 
 \be
 \sigma(t) = M_P\sqrt{\frac{n(n+2)}{2}}
 \ln\left(\frac{b}{b_0}\right)~.
 \label{radion}
 \ee 
This field is usually called the radion, 
and it is defined in such a way that it sets to zero
when the stabilized volume is reached.
Indeed, with the use of this radion, the last effective 4D action becomes
 \be
 S =  \frac{M_P^2}{2} \int\!d^4x\sqrt{|g_{(4)}|}\,  R_{(4)}
  + \,\frac{1}{2} \int\!d^4x \sqrt{|g_{(4)}|}\,\left(\partial^\mu \sigma\right)
  \left(\partial_\mu \sigma\right)~;
 \label{4ds}
 \ee
where the last term  corresponds to the action of a run
away scalar mode. Hence, without potential, the radion field can take
any value. Furthermore,  under any perturbation, the volume of the
extra space is totally unstable. In general, an active radion means
a variable bulk, and it could be seen as an unwanted and harmful scenario. 
As we will discuss below, this field couples to all other fields in the
theory, affecting dispersion relations and the definition of coupling
constants. Also, its couplings to the inflaton may introduce potential
threats to standard cosmology (see for instance Refs.~\cite{mmp}).
This can be disastrous, and thus, it is important to provide a radion
potential capable to keep the radion at its zero value.

Some comments are in order. The very definition of the radion depends on the
background metric we have chosen. Different geometries would mean  different
mathematical forms for the radion field, but the last would always be present. 
Flat backgrounds are the simplest examples where 
calculations can be worked out very clearly. 
Thus, hereafter we will assume the bulk to be flat.
Nevertheless, one has
to keep in mind that, in any realistic scenario, backreactions
due to the energy that sources the stabilization potential may require to
refine the compactification analysis to take such effects into account. 
To keep our analysis simple, however, we will neglect such effects. 

As already mentioned, in what follows, 
we shall consider two possible sources of energy contributing to stabilization.
First,  pure cosmological constants, 
which are usually seen as the zero level energy  
produced by the actual physics living on the
space-time of the theory. Actual cosmological constant is rather
small and one can safetly take it to be zero for simplicity, 
but in a theory with extra dimensions what we see in four dimensions,
is just the result of
all contributions that come from the various sectors of the theory. Thus, in
a bulk-brane scenario, both bulk and brane cosmological constants could be
expected. 
Branes, of course, should be located
at the fixed points on the compact dimension. 
Next possible source comes from position dependent vacuum configurations on the
bulk, which can be model using bulk scalar fields.
These could
actually come from many sectors of string theory, usually as extra degrees of
freedom of vectorlike o tensorlike  fields. However, 
we will not need to make any assumption on the origin of such
fields other than they having non trivial bulk configurations. That will allow
us
to keep our analysis general,  and to address the question of whether such
ingredients could be enough to built appropriate stabilization potentials 
from a  phenomenological perspective.

%%%%%%%%%%%%%%%%%%%%%

\subsection{Radion couplings and effective potentials}

Before entering into the discussion of the stabilization mechanism,
we shall first make a note on the effect of the above introduced
conformal transformation [Eq. (\ref{conformal})]
on other physical actions besides that of gravity.
Consider for instance a  bulk scalar
field, $\phi(x,y)$. The corresponding action,
in the initial $(4+n)$D  frame,
before performing the conformal transformation on the metric, goes as
 \be
 S_{\phi} = \int\!d^4x\,d^ny\,\sqrt{|g_{(4)}|}\,\sqrt{|h|}
 \left[\frac{1}{2}G^{AB}\partial_A\phi\,\partial_B\phi
 - U(\phi)\right]~.
 \label{sphi}
 \ee
Without lost of generality, we  can always assume that $\phi$ has a
proper Kaluza Klein (KK) mode decomposition, which should be defined
for each given topology of the compact space. Such modes are in
general the orthogonal solutions to the  free equation of motion,
only considering upto mass terms in the above general
action,
with the proper boundary conditions. A typical expansion should have
the formal expression
 \be
 \phi(x,y) =
 \sum_{\vec{n}}\frac{\xi_{\vec n}(y)}{\sqrt{vol_n}}\,\phi_{\vec{n}}\,(x)~;
 \ee
where $\vec n$ stands for all the KK indices; and the KK modes, $\xi$, should
obey the formal normalization condition
 \be
 \int\! d^ny\,\xi_{\vec{n}}\,\xi_{\vec{n}'}= \delta_{\vec{n}\,\vec{n}'}~.
 \ee
By introducing this expressions in the action, and including the
conformal transformation, we get
 \be
 S_{\phi} = \int\!d^4x\, \sqrt{|g_{(4)}|}
 \left[\sum_{\vec{n}} \left(\frac{1}{2}g^{\mu\nu}
 \partial_\mu\phi_{\vec{n}}\,\partial_\nu\phi_{\vec{n}}\right)
  - e^{-\alpha\,\sigma/M_P}\, U_{eff}(\phi_{\vec{n}})\right]~;
 \label{sphic}
 \ee
where  we have replaced the conformal factor terms in favor of the
radion field, explicitly using the equivalent expressions
 \be
 e^{2\varphi} =\left(\frac{vol_n}{\sqrt{|h|}~}\right)=
  \left(\frac{b_0}{b}\right)^n = e^{-\alpha\,\sigma/M_P}
 \ee
with  $\alpha = \sqrt{2n/(n+2)}$~. Thus, we notice that in the
conformal frame the radion couples exponentially to an effective
potential, which is  formally defined through the integral
 \be
 U_{eff} =vol_n\cdot\int\!d^ny\,
 \left(\frac{1}{2}\, \frac{\vec{\nabla}_y\phi\cdot\vec{\nabla}_y\phi}{b^2}
 +  U(\phi)\right)~,
 \ee
with $\vec\nabla_y$ the gradient on the compact space  coordinates. 
Note also that last expression actually 
corresponds to the potential energy, $U_{ini}$, 
one  calculates in the initial frame, 
but for the global factor $vol_n$ instead of the physical volume
$\sqrt{|h|}$. In fact, 
one can also writes $U_{eff}=e^{-\alpha\,\sigma/M_P}U_{ini}$.
The first term in above equation would contribute to the whole potential in 
Eq.~(\ref{sphic}) with the KK mass term. The KK squared mass, as usual, appears
proportional to the squared inverse physical radius, $b^{-2}$, 
up to an overall
conformal factor $(b_0/b)^n$. So, in Einstein frame, the effective mass of KK
modes should  follow the time dependence of radius variations with a power law
modulation.

The overall conformal factor on  potential terms, is in fact a general
feature for most actions. It also appears, for instance, in the case of a
bulk cosmological constant, where the action
$S_\Lambda = \int\!d^4x \,d^ny\, \sqrt{|g_{(4+n)}|} ~\Lambda$~,
is easily integrated over the extra dimensions, to become in Einstein
frame
 \be S_\Lambda = \int\! d^4x
  \sqrt{|g_{(4)}|}~\Lambda_{n}\,e^{-\alpha\,\sigma/M_P}~,
  \label{slambda}
  \ee
with the effective cosmological constant $\Lambda_n =
vol_{n}\cdot\Lambda$. We must take into account these overall factors
when discussing any bulk generated potential.

Similarly, for brane fields, although 4D actions do not involve
a $\sqrt{|h|}$ factor but just
$\sqrt{|g_{(4)}|}$ for the induced metric on the brane,
the conformal transformation we used for
the rest of the theory shall introduce a radion coupling at least for brane
scalar and fermion fields as well as the brane cosmological constant. 
Indeed, for a scalar field, $\chi$, one obtains
 \be
  \int\! d^4x\, \sqrt{|g_{(4)}|} \, e^{-\alpha\,\sigma/M_P}
  \left( \frac{1}{2}g^{\mu\nu} \partial_\mu\chi\,\partial_\nu\chi -
  e^{-\alpha\,\sigma/M_P} U(\chi)\right)~,
 \ee
whereas one gets
 \be
 \int\! d^4x\, \sqrt{|g_{(4)}|} ~ e^{-\frac{3}{2}\alpha\,\sigma/M_P}
 ~g^{\mu\nu}e^a_\nu\left(i\bar\psi D_\mu\,\gamma_a\psi\right)~,
 \ee
for a massless brane fermion, $\psi$.
In the case of a 3-brane cosmological constant, $\lambda$, one has
 \be 
 \int\! d^4x\, \sqrt{|g_{(4)}|} \, e^{-2\alpha\,\sigma/M_P}\, \lambda~.
 \label{lambda}
 \ee
 There is no coupling of
radion field to massless gauge fields, though. Above couplings affect
the definition of canonical brane field through  the modification of
the kinetic terms, and hence the corresponding dispersion relations.
They also transform cosmological constants into radius functions.
Clearly, all this results reduce to  usual ones for a stable bulk,
when $b=b_0$ ($\sigma=0$), and certainly,  one can use
standard expressions at first order, when the radion is close to the
minimum, such that its couplings can be treated perturbatively.

%%%%%%%%%%%%%%%%%%%%%%%%%%%%%%%%%%%%%%%%%%

\section{Radion stabilization by vacuum energy}

 As we already mentioned, some ideas on how to generate an
stabilization potential for the radion can be found already in the
literature~\cite{kaluza-klein,tsujikawa,kklt,Frey,Joe1,quiros,wise,maru,chacko}. In
particular, for a single extra dimension, it is has been pointed
out~\cite{wise,chacko} that  a radion  potential can be produced if
translational invariance is broken in the bulk by the vacuum
expectation value of a scalar field. Here, we will further explore
this idea for flat extra dimensions. We shall perform our analysis in the 
conformal frame where the radion has been identified.
The basics of the mechanism we are exploring are rather
simple. Bulk field configurations may provide an effective 4D energy.
Furthermore, if bulk energy density breaks translational invariance
along the bulk coordinates, one gets different amounts of energy for
different volume sizes, thus, generating a potential energy,
which we shall find
convenient  to write in terms of the radius, as
$U_{rad}(b)$. Of course, if there is a non trivial minimum for
$U_{rad}(b)$, this would be identified as $b_0$.

%%%%%%%%%%%%%%%%%%%%%%%%%%%%
\subsection{Stabilization by cosmological constants}

Lets us first notice that the use of only a cosmological constant, either from
3D-branes located at fixed points or the bulk, does not provide a desirable
scenario. For any individual case the radion potential is just an exponentially
decaying function without
non trivial minimum. However, the
combination of both contributions 
may work. From
Eqs.~(\ref{slambda}) and (\ref{lambda}),  
the most general radion potential one can build in this case is 
 \be 
 U^\lambda_{rad}(\sigma) = e^{-\alpha\,\sigma/M_P}
 \left(\Lambda_n + e^{-\alpha\,\sigma/M_P}\,\lambda\right)~.
 \label{ucc}
 \ee
Clearly $U^\lambda_{rad}(0) = \Lambda_n +\lambda$, whereas 
$U^\lambda_{rad}\rightarrow 0$ for 
$\sigma\rightarrow \infty$. This potential has a minimum at 
$\sigma_0 = (M_P/\alpha)\,\ln(-2\lambda/\Lambda_n)$, provided $\lambda>0$. 
The requirement that $\sigma_0=0$ be a minimum
implies that $\Lambda_n +2\lambda =0$. At first sight this condition might be
seen as a fine tuning, nevertheless,  
this is  actually  what fixes the stable radius to 
$b_0 = (-2\lambda/\Lambda)^{1/n}$. 
Therefore, the only 
appropriate
potential for this  case goes as
 \be 
 U^\lambda_{rad}(\sigma) = \lambda\,e^{-\alpha\,\sigma/M_P}
 \left( e^{-\alpha\,\sigma/M_P}-2\right)~.
 \label{ulambda}
 \ee

%\vskip1em
%%%%%%%%%%%%%%%%%%%%%
\begin{figure}%[ht]
\centerline{
\epsfxsize=220pt
\epsfbox{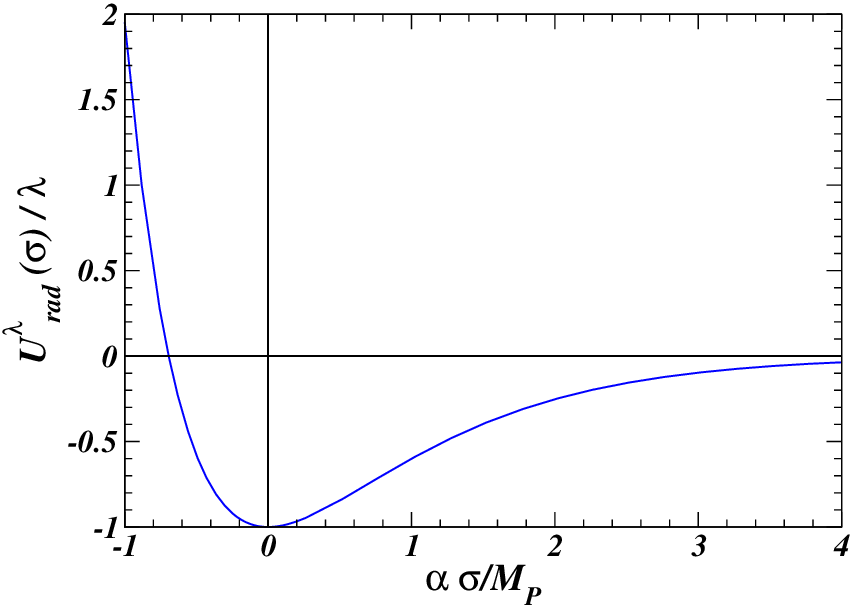}
}
%\vskip1ex
\caption{Radion stabilization 
potential profile generated by the sole introduction of brane and bulk
cosmological constants according to Eq.~(\ref{ulambda}).}
\end{figure}
%%%%%%%%%%%%%%%%%%%
%\vskip2em

In figure 1 we have plotted the general profile for this potential. 
It is worth noticing that the potential diverges exponentially 
for negative values of $\sigma$, or radii smaller than $b_0$. This suggest that
our configuration of bulk and brane cosmological constants works perfect to keep
the extra dimension from collapsing into itself. However,  
as it is
clear, the depth of the potential is given by a single parameter
which by construction can not be fixed to zero,
the brane cosmological constant $\lambda$.  
This implies, at the stable volume configuration, a non zero  and negative
effective 4D cosmological constant, $U^\lambda_{rad}(0) = -\lambda$. 
This is troublesome since the observed
cosmological constant is rather small and positive.
 Furthermore,  from 
the potential one gets a Planck suppressed 
effective radion mass at the minimum, 
 \be m_\sigma = \alpha \sqrt{2\lambda}/M_P~,
 \label{ms0}
 \ee 
which may also imply a too light radion mass, against observational limits
on gravity strength coupled scalars, that indicate $m_\sigma>10^{-3}~eV$, which
would require that $\lambda> TeV^4$.
These features 
are indeed a potential problem, and thus, 
one is force to departure from this 
simple model. 
We should notice that  
same conclusions are reached when one does the 
analysis for the radius instead of  the radion field, as expected.

It is worth stressing  the fact that the potential naively
calculated in the initial frame, that is with out properly 
including the conformal factors, 
is only a polynomial function of the radius,
$b^n\Lambda+\lambda$, whose only minimum at the best
resides at $b=0$.   
This clearly shows the
risk of getting misleading results when the analysis 
is not properly performed in the Einstein frame, 
and the conformal factors are taken into account.

An alternative for $n>1$ could be to add brane tensions 
at the natural boundaries of the compact space too. 
A realization of such an scenario from string theory
may of course need the introduction of 
intersecting brane configurations.
For instance, 
a $(n+2)$-brane tension, $\tau$, would 
contribute to the effective action with the term
 \be
  \int\! d^4x\, \sqrt{|g_{(4)}|} \, e^{-\beta\,\sigma/M_P}\, \tau_n~,
 \label{tau}
 \ee
where $\beta=(n+1)\alpha/n$ and $\tau_n=b_0^{n-1}\tau$. Now, 
$ U_{rad}^\tau=
e^{-\alpha\,\sigma/M_P}(\Lambda_n + e^{-\alpha\,\sigma/M_P}\,\lambda) 
  + e^{-\beta\,\sigma/M_P}\,\tau_n$~, 
has a minimum for $\sigma=0$ provided 
$\alpha(\Lambda_n + 2\lambda)+\beta\tau_n=0$. An additional condition can now be
imposed by requiring that $U_{rad}^\tau(\sigma=0)=0$, which gives 
$\Lambda_n + \lambda +\tau_n=0$. This condition   would always imply
that at least one of the cosmological constants we are considering is negative,
and conspire to (almost) cancel the effective cosmological constant.
 By combining these two equations we find the
unique solution $\Lambda_n = (n-1)\lambda$, that fixes the radius at 
$b_0 =[(n-1)\lambda/\Lambda]^{1/n}$, and also $\tau_n = -n\lambda$. 
Thus, the potential becomes 
 \be 
 U_{rad}^\tau= \lambda\,e^{-\alpha\,\sigma/M_P}
 \left[n\left(1 -e^{-\alpha\,\sigma/nM_P}\right) +
 \left(e^{-\alpha\,\sigma/M_P}-1\right) \right]~,
 \label{utau}
 \ee
whereas the associated radion mass is now
 \be 
 m_\sigma^2=\left(\frac{n-1}{n}\right)\frac{\alpha^2}{M_P^2}\,\lambda~.
 \ee
Note that last expression
has a similar form than the result given above [Eq. (\ref{ms0})]. Again, it
now implies that all cosmological constants  in the model obey 
$\lambda,\sim\Lambda_n,\sim|\tau_n|> TeV^4$.
Notice that $U_{rad}^\tau$ decays exponentially for large $\sigma$. Thus, the
potential profile presents a potential barrier that isolates the local minimum
$\sigma=0$ from infinity, as it is depicted in figure 2. This is going
to be a constant feature for the examples we shall discuss below. This, of
course, may indicate the risk of a possible spontaneous 
decompactification by quantum 
tunneling. However, notice that the 
width is given in
Planck mass 
units, above which we can not really trust our effective analysis due
to possible stringy (or quantum gravity)
effects, that we have not under control in here, 
and that could substantially modify the potential  for
larger values of $\sigma$. This issue 
is out of the scope of the present paper, and 
therefore, we will not discuss it any further beyond 
this note,
neither we will do for the cases presented below, when it appears.

%\vskip1em
%%%%%%%%%%%%%%%%%%%%%
\begin{figure}%[ht]
\centerline{
\epsfxsize=220pt
\epsfbox{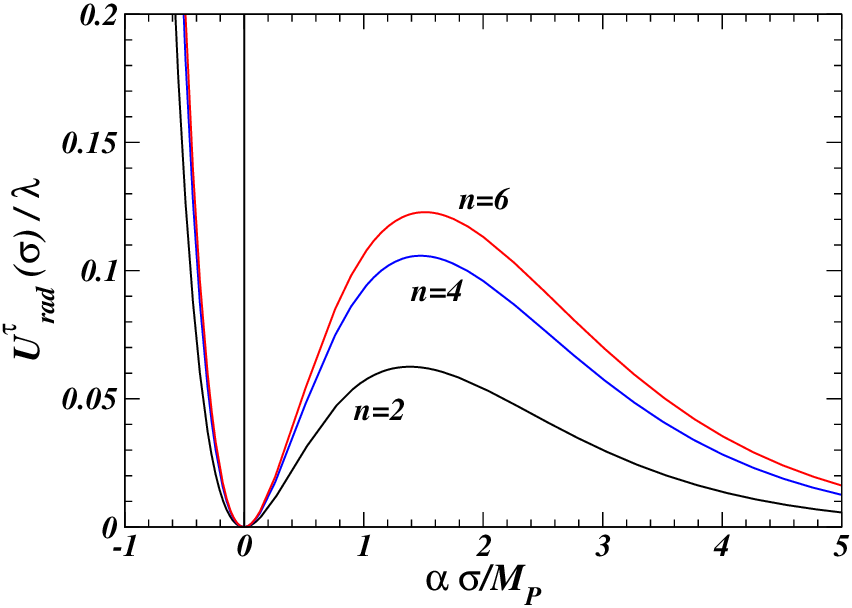}
}
%\vskip1ex
\caption{Radion potential profile generated by 
cosmological constants according to Eq.~(\ref{utau}), for $n$ as indicated.}
\end{figure}
%%%%%%%%%%%%%%%%%%%
%\vskip2em
%%%%%%%%%%%%%%%%%%%%%%%%%%%%%%%%%%%%

\subsection{Potential building}

Next simplest example one can provide for  bulk energy is a
$y$-dependent vacuum.   That arises in models where non trivial boundary
conditions are imposed on a bulk scalar field configuration. To
elaborate,  let us consider a massive scalar field, $\phi$, described
by the action given in Eq.~(\ref{sphi}) for $U(\phi) = \frac{1}{2}m^2\phi^2$. 
Therefore, the vacuum
configuration in the initial frame, with a given volume of size $b$,
should be a solution to the equation of motion
 \begin{equation}
 \label{ec:mov}
 \left[-\nabla_y^2 + \kappa^2\right]\langle\phi\rangle(y)=0~,
 \end{equation}
where $\kappa=mb$ 
and $\nabla_y^2$ is the Laplacian operator on the extra dimension
coordinates. Notice also that we are considering only
those vacuums which do not break translational invariance along the
brane. Above equation should be complemented with the boundary
conditions defined at the end points of the compact manifold. Without
them there would be no y-dependent vacuum energy in the minimal configuration. 
On the orbifold, for instance, these conditions are given on
3-branes located at the fixed points. They  define localized
sources for the bulk vacuum. These boundary conditions
may be due to some other physics sited on the branes which forces the
bulk field to pick up a non trivial vacuum expectation value. As an
example one can consider the coupling to some brane scalar field, $\chi$, as
$\chi\phi\delta(y-y_0)$, where $y_0$ is a fixed point where the brane is
located. If $\chi$  develops a vacuum expectation value (vev), this shall
induce a vev on $\phi$ that varies along the bulk. This mechanism
has been used in several models to explain small vevs in
distant branes~\cite{shiny}. Another possibility may be the existence of
localized potential terms for the bulk field  on the branes, which favors
a non trivial localized vev. This happens, for instance, if one considers
a Higgs type localized potential $(\phi^2- v)^2\,\delta(y-y_0)$.

Regardless the mechanism that fixes the boundary conditions for
$\langle\phi\rangle$, these induce a non trivial profile for the vev
along the bulk. Once such a vev is given,
by setting it back into the Lagrangian, ${\cal L}$, at any given radius $b$,
one formally gets   in the Einstein frame
the radion potential contribution:
 \be
 U^\phi_{rad}(b)=\left(\frac{b_0}{b}\right)^{2n}\,U_{ini}(b) ~,
 \label{uradphi}
 \ee
here written in terms of the radius, and where we have conveniently used
the potential as it is read in the initial frame (before conformal
transformation), 
\begin{equation} 
 U_{ini}(b)=-\,b^n\cdot\int\! d^{n}y ~\cal{L(\langle\phi\rangle)}~.
 \label{uini}
 \end{equation}
By written the potential
this way, it becomes clear that in general a minimum for 
$U_{ini}$ is not a
minimum of $U^\phi_{rad}$. 
The conformal factor deforms the potential, and may even 
compromise stabilization in some cases. 
Nevertheless, the function $U_{ini}$  
will proof to be a useful reference when 
analizing the radion potential, as we shall see in the examples below.

On the other hand, even if one has a non trivial minimum for above potential, 
there is no guarantee that the potential would be
zero at minimum. Such a case can, however, be controlled with 
the addition of cosmological constants. Therefore, the generic potential
one can build  goes as
 \be
 U_{rad}(b) =\left(\frac{b_0}{b}\right)^{n} 
            \left[\left(\frac{b_0}{b}\right)^{n}
	    \left[U_{ini}(b) + \lambda\right] +\Lambda_n\right]~.
 \label{urad}
 \ee
It is worth noticing that the explicit $b_0$ dependence  
on above equation, although it may be annoying, 
it is actually harmless. It is 
introduced by the conformal
factors, and we keep it everywhere 
just for dimensional reasons. Nevertheless, 
it is actually ignored for the minimization 
of the potentials, since 
we can always factor it away  using that $\Lambda_n=b_0^n\,\Lambda$.

Before working out some examples, we notice 
that the radion mass
provided by our mechanism has the general form
 \be
 m_\sigma^2 = \left(\frac{b_0\alpha}{nM_P}\right)^2\,
 \left[\frac{d^2U_{rad}(b)}{db^2}\right]_{b=b_0}~.
 \label{msigma}
 \ee
Therefore, it always come with a Planck suppression, which may, 
however, be overcomed provided the potential well is steep enough.

Among the many situations in which a minimum could appear for the radion
potential in Eq.~(\ref{urad}), two are of special interest for model
building. Both are realized when $U_{ini}(b)$ has already 
a non trivial minimum, $b_{i}$.
 \begin{itemize}
 \item[(i)] First, one
 can always guarantee that the actual minimum in the conformal frame remains 
 the same, such that $b_0=b_{i}$. This actually happens when the brane
  cosmological constant is used to shift the minimum of $U_{ini}(b)$ to zero in 
  the initial frame,  by  choosing  $\lambda = -U_{ini}(b_{0})$, and one takes
  $\Lambda =0$ to insure a zero effective cosmological constant in the conformal
  frame.
   In this case, the overall power law factor in Eq. (\ref{urad}) 
  has no impact on the  location of  the minimum of the potential. 
  As a matter of fact, it is easy to see that  within this conditions
  $U_{rad}(b_0) =0$, whereas  
 $U_{rad}'(b_0)=U_{ini}'(b_0)-\frac{2n}{b_0}\left[U_{ini}(b_0)+\lambda\right]=0$;  
 and $U_{rad}''(b_0)=U_{ini}''(b_0)$, from a similar reasoning.
 Last equation also shows that 
 the radion mass can be calculated directly from $U_{ini}(b)$.
 Hence, it is usually enough to stablish the existence 
 for a non trivial minimum on $U_{ini}(b)$ to know that there is a working
 situation in the Einstein frame.

 \item[(ii)] Second, one can take advantage of the interplay among 
 the two cosmological constants to provide more
 control on potential depth. Key observation is that for any given function,
 $f(b)$, its zeros are fixed points under the modulation by a $1/b^n$ factor,
 provided $b\neq 0$.
 This is actually the analytical reason why the minimum of $U_{ini}(b)$
 is kept in previous situation despite the conformal factors. 
 Also, since $1/b^n$ is always positive, 
 it does not change the sign of any given value of $f(b)$. However, it
 suppresses the function for large $b$. 
 Thus, one can subtract a large cosmological
 constant, $\lambda\ll -\|U_{ini}(b_{i})\|$; to $U_{ini}(b)$,
 to shift the minimum towards large negatives values, 
 as to compensate for the $1/b^n$ modulation, and provide a
 deeper well for the effective potential:
 \be 
 U_{eff}(b) =\left(\frac{b_0}{b}\right)^{n}
	    \left[U_{ini}(b) + \lambda\right]~.
 \label{ubeff} 
 \ee
 Finally, a large positive $\Lambda_n = -U_{eff}(b_0)$ should be chosen in order to 
 cancel the radion potential at the minimum. It is not hard to see that in 
 in this scenario  $b_0\neq b_{i}$. As a matter of fact, the minimum of
 $U_{eff}$ shall now also 
 become the minimum of the above radion  potential (\ref{urad}). Moreover, since
 also  $U_{rad}''(b_0) = U_{eff}''(b_0)$, the radion mass may, in this case, be
 calculated directly from the effective potential instead. 
\end{itemize}

It is not difficult to understand what a mismatching
$\delta\lambda= U_{ini}(b_i)+\lambda\neq 0$ 
does for the deviation of the actual minimum, $b_0$, from  $b_i$
in previous scenarios.
We can imagine a simple situation where $\delta\lambda$ is small, such that in
the limit where it is neglected we start with the minimum at $b_i$, as 
described
in the first item above.
By switching on $\delta\lambda$ we shall be moving into the
second scenario just described. So, the actual minimum should now 
be displaced from $b_i$  by $\delta b=b_0-b_i$. Being the
minimum of $U_{eff}$, $b_0$ 
fulfills the condition $b_0\,U'_{ini}(b_0)-n\delta\lambda=0$, 
which at first order gives
 \be 
 \delta b\approx \left(\frac{n}{b_iU_{ini}''(b_i)}\right)\delta\lambda~.
\ee
As the  coefficient within parenthesis is positive by definition, we conclude
that the  minimum is shifted according to the sign of $\delta\lambda$,
and clearly, we  require $\Lambda_n\approx -\delta\lambda$.
On the other hand, by looking at the second derivatives of the potentials, 
we find that at $b_0$ one gets $U_{eff}'' = U_{ini}'' +\delta U''$, where
 \be
 \delta U'' \approx -
  \frac{n(n+1)}{b_i^2}\,\delta\lambda~.
 \ee
Therefore, for $\delta\lambda<0$, the potential around the minimum gets 
tighten and the radion mass increased.

The case where  the $U_{ini}$ minimum is trivial, 
meaning $b_i=0$ or infinity, is hard to 
handle in general.  However, as 
in the case of sole cosmological constants, there may be some scenarios where
$U_{rad}$ do have a non trivial minimum. Whether this is so would have 
to be studied for each particular case, though.
We will illustrate this situations along next section.

\section{Radion stabilization on orbifolds}
%%%%%%%%%%%%%%%%%%%%%%%%%%
\subsection{The interval}

To exemplify the mechanism let us elaborate on the simplest case of one single
extra dimension where the coordinate $y$ takes values in the interval $[0,1]$.
The general solution to the equation for the vacuum state (\ref{ec:mov}) 
is then
\begin{equation}
 \label{general}
 \phi(y)= A e^{\kappa y}+Be^{-\kappa y},
 \end{equation}
where the constants $A$ and $B$ are given in terms of the boundary conditions,
which we assumed to be $\phi(0) = v_0$ and $\phi(1) = v_1$, where $v_{0,1}$ have
mass dimension $3/2$ by definition. Thus, one gets
 \begin{equation}
 A = \frac{v_1-v_0 e^{-\kappa}}{ e^{\kappa}- e^{-\kappa} }~;
 \qquad \mbox{and} \qquad
 B = \frac{v_0 e^{\kappa}-v_1 }{ e^{\kappa}- e^{-\kappa} }~.
 \end{equation}
It is straightforward to calculate the  potential in the initial frame 
according to Eq.~(\ref{uini}), which goes as  
  \begin{equation}
  \label{veff}
  U_{ini}(b)= \frac{m}{2}
  \frac{\left(v_0^2+v_1^2\right)\,\cosh\, \kappa -2v_0v_1}{\sinh\, \kappa}.
  \end{equation}
Note that the potential is invariant under the exchange 
$v_0\leftrightarrow v_1$. This was expected because the physical 
situation we are describing within  the interval 
(equivalent to the one dimensional orbifold $S^1/Z_2$)
is invariant under exchange of the boundaries, 
which can be seen as an effect of parity symmetry.
This potential has a sizable minimum at 
  \begin{equation}
 mb_{i}=arccosh \left(\frac{v_0^2+v_1^2}{2v_0v_1}\right)~.
 \label{mbi}
 \end{equation}
Hence, the stable radius is proportional to the inverse mass of the bulk 
scalar field by a factor fixed by the boundary conditions, which ranges from
zero to infinity. This provides a great freedom on the bulk scalar mass, and
allows for a simple realization of large extra dimensions, at the price of 
moving the hierarchy to the boundary conditions.  Particularly, for large 
$v_0/v_1$ ratios 
one gets the approximate expression $mb_0\approx [\ln(v_0/v_1)]^2$.
At the minimum we get
 \be 
 U_{ini}(b_{i}) = \frac{m}{2}\|v_0^2-v_1^2\|~,
 \label{uinimin}
 \ee
and so the potential is always positive.
Notice also that the potential goes asymptotically to a constant:
$U_{ini}(b\rightarrow \infty)= m(v_0^2 + v_1^2)/2$, 
and for small $b$
behaves like $\sim (v_0-v_1)^2/2b$, provided $b_0\neq 0$. 

Clearly, $v_0=v_1$
is not a favored scenario. First of all, it implies $b_{i}=0$, where the
potential vanishes. Nevertheless,  
the $1/b^2$ squared modulation removes this minimum and kills the asymptotic
behavior, such that the only possible minimum in the Einstein frame
becomes $b\rightarrow \infty$.
Furthermore, by including brane and bulk 
cosmological constants one  gains new terms that go as
$\Lambda/b + \lambda/b^2$. As we have shown, this piece of the potential has a
non trivial minimum by itself, provided $\lambda >0$ and $\Lambda<0$. Same
situation holds for the whole radion potential.  This can be easily seen 
as follows. First, consider that close to zero $U_{rad}(b)$ diverges as
$\lambda/b^2$. Thus $b\neq 0$ requires a positive $\lambda$. 
Next, we notice that $\Lambda$ has to be negative to compensate 
the other monotonic and positive defined parts
of the potential to provide a minimum. However, we now notice that the 
asymptotic form for the potential goes as $\sim\Lambda/b$, and thus $U_{rad}(b)$
reaches zero asymptotically from below, which implies that $U_{rad}(b_0)$ is
strictly negative. Therefore, we are drove to this conclusion: 
one can find a way to provide a stabilization
potential in this case, but one always ends with a
non zero cosmological constant,
which is not very attractive.
This suggests that asymmetric
boundary conditions on both ends of the interval may be preferred. 
Notice, however, that for either $v_0$ or $v_1$ null, 
$b_0$ would go to infinity, and we will end in a similar situation. 
 
Next, we  proceed to study the radion potential in the Einstein frame by
assuming that $v_0> v_1$, for simplicity. 
The opposite case is actually equivalent due to the 
$v_0\leftrightarrow v_1$ exchange symmetry. 
As the potential in the initial frame has already a minimum, the two
scenarios for model building described in previous section shall be  useful. 

As the first approximation we add a brane 
cosmological constant $\lambda = -m(v_0^2 - v_1^2)/2$, and take $\Lambda=0$. 
Thus, the resulting radion potential, 
$U_{rad}(b)= (b_0/b)^2[U_{ini}(b)+\lambda]$, keeps the minimum at $b_0 =b_i$,
as defined by Eq.~(\ref{mbi}), and fixes $U_{rad}(b_0)$ to zero. 
However, now $U_{rad}(b)$ approaches zero
asymptotically  like $\sim mb_0^2v_1^2/b^2$ for large $b$, and so, 
an infinite $b$ also appears as a possible stable configuration.
Both minima are separated each other by a  potential
barrier, and so there is the slight possibility of tunneling for the radion
when perturbed. Of course, 
the high and the width of the potential barrier depends on the
parameters of the theory, particularly on the size of the boundary conditions, 
and one may hope some 
configurations with large values for $mv_1^2$, 
would ameliorate this possible
trouble. All this features can be observed 
in figure 3 where we have plotted this radion potential (continuous lines) in
units of  $mv_1^2/2$ to make it dimensionless, for different values of 
the  $v_0/v_1 $ ratio. 
Notice that, as expected,  
a larger $v_0/v_1$ ratio tends to increase the relative
height and width of the potential barrier, making the potential well deeper and
narrower. And, at the same time, rising  the hierarchy among $b_0$ and
$m$. 

%%%%%%%%%%%%%%%%%%%%%
\begin{figure}%[ht]
\centerline{
\epsfxsize=220pt
\epsfbox{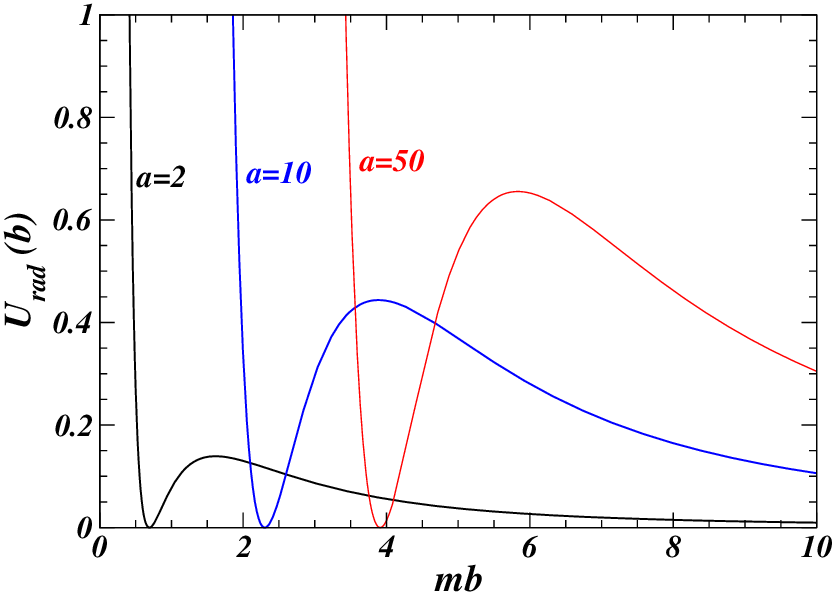}
}
%\vskip1ex
\caption{Radion stabilization 
potential profiles generated by vacuum energy in the interval, in unites of
$mv_1^2/2$, for given values of 
$a = v_0/v_1$, as indicated. }
\end{figure}
%%%%%%%%%%%%%%%%%%%
%\vskip2em

Using Eq.~(\ref{msigma}) we calculate the radion mass for this case and obtain
 \be
 m^2_{\sigma} = \frac{4}{3} \left(\frac{m}{M_P}\right)^2  
 \frac{v_0^2v_1^2}{m\|v_0^2-v_1^2\|}
 \left[arccosh \left(\frac{v_0^2+v_1^2}{2v_0v_1}\right)\right]^2~.
 \ee
Thus, the radion mass is  also sizable by adjusting the boundary conditions,
just as the size of the extra dimensions is so, according to Eq.~(\ref{mbi}).
For a large $v_0/v_1$ ratio above equation can be approximated as
$m^2_\sigma\approx (4/3)[\ln(v_0/v_1)]^2\, mv_1^2/M_P^2$, which means that if
$\ln(v_0/v_1)\sim O(1)$, then 
$mv_1^2>TeV^4$ 
to maintain $m_\sigma>10^{-3}~eV$, but this also would indicate that $b_0$
cannot be too large.  On the contrary, a larger hierarchy would
easily provide a large radion mass, without implying a large $m$, and so
allowing for larger compactification radius.

In a second approach,  one can use the cosmological constants to
greatly improve on the potential depth, as described in previous section.
Notice, however, that this procedure will not substantially change 
the asymptotic 
behavior of the radion potential, because we shall only choose a different 
set of cosmological constants,  keeping the functional form of the
potential as given in Eq.~(\ref{urad}) with $U_{ini}$ replaced by 
Eq.~(\ref{veff}). Yet, for large $b$ we get, 
$U_{rad}(b)\sim b_0^2\Lambda/b$, where  now the chosen  
$\Lambda= -U_{eff}(b_0)$
could actually become quite large.
Thus, the radion potential shall remain with 
two local minima, $b_0$ and infinity, but now with a wider and taller 
potential barrier in between.
The corresponding effective potential
do have a non trivial minimum, as the non zero solution for $\kappa= mb$
in the equation
\[4\lambda\sinh\kappa + 
 m\left(v_0^2 + v_1^2\right)\left(\sinh 2\kappa + 2\kappa\right)
 = 4mv_0v_1\left(\sinh \kappa + \kappa\cosh\kappa\right)~.\]
There is no analytical solution to last expression, and  thus,
one has to proceed numerically in most cases, or at least perturbativelly 
for small displacements. As discussed already, since we are now using 
$\delta\lambda<0$, we can expect a minimum  shifted to smaller
values, and a tighter potential well for larger values of $|\lambda|$. 
All this is confirmed by the numerical analysis, as it can be checked in
figure 4, where  we have plotted the radion potential profile for
the ratio $v_0/v_1 = 50$, and for some different values 
of $\lambda$, chosen as numerical multiples of $U_{ini}(b\rightarrow\infty)$
as for example.

%%%%%%%%%%%%%%%%%%%%%
\begin{figure}%[ht]
%\vskip2em
\centerline{
\epsfxsize=220pt
\epsfbox{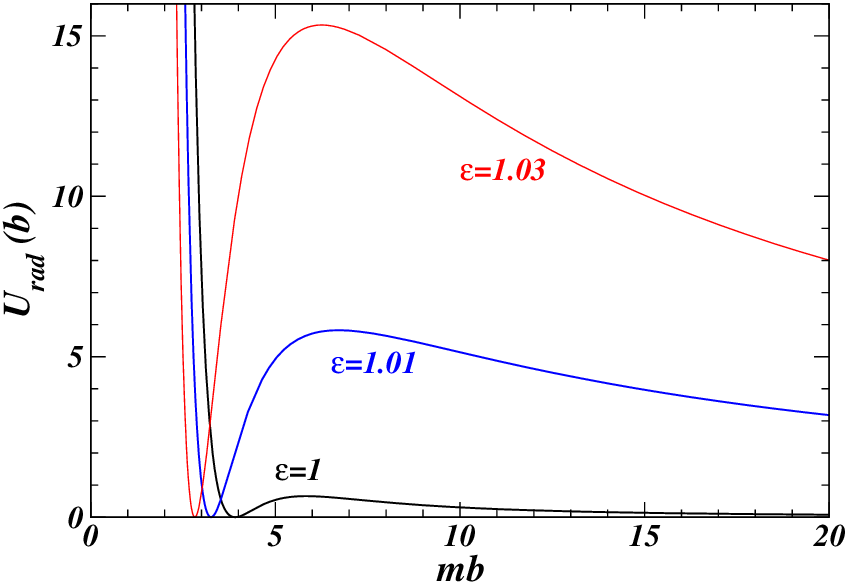}
}
%\vskip1ex
\caption{Radion stabilization 
potentials for one extra dimension, with non zero $\Lambda$. 
As before,  potentials are plotted in units of $mv_1^2/2$, 
for the ratio $v_0/v_1=50$ and for slightly different values of 
$\lambda = -\varepsilon (v_0^2/v_1^2-1)$  
in same units as the potential, with $\varepsilon$ as indicated. }
\end{figure}
%%%%%%%%%%%%%%%%%%%
%\vskip2em
%%%%%%%%%%%%%%%%%%%%%%%%%%

%%%%%%%%%%%%%%%%%%%%%%%%%%%%%%%%%%%%%%%
\subsection{The $T^n/Z_2$ orbifold}

Let us now explore in some detail a more general example.
Next, we shall consider a model where the bulk manifold is given by a $T^n/Z_2$
orbifold, where the $Z_2$ corresponds to the identification of points  
on the symmetric $T^n$ torus with common radii $b$, according to the mapping 
$\vec{y} \rightarrow -\vec{y}$. 
To simplify matters, we will consider
only the whole volume variations  which do not alter this overall geometry,
such that the metric on the compact space remains of the form 
$ds^2_{compact}= b^2\delta_{ij} dy^idy^j$, where the $y_i$ coordinates on the
torus have values in the interval ${\cal I} = [-1,1]$. Of course, on the 
orbifold, physical compact space is smaller. It can be chosen to be
represented by the reduced  
$[0,1]\times {\cal I}\times\cdots\times{\cal I}$ space.
In this orbifold, there are  
$2^n$  fixed points which correspond to the vertices of 
the unitary hypercube  
${\cal I}_0^n={\cal I}_0\times\cdots\times{\cal I}_0 $, 
where ${\cal I}_0 = [0,1]$.
This symmetric $T^n/Z_2$ orbifold  
has a residual discrete symmetry ${\cal R}_{\pi/2}^n$,
given as the set of
rotations by $\pi/2$ around any $y_i$ coordinate axis.
This symmetry transformations  map
fixed points, located at the same distant from the origin, among themselves.    

Since the potential we are to build is due to  boundary conditions 
on the fixed points, the fact that
all $y$ directions should have the same size suggests
a totally symmetric potential
under the same ${\cal R}_{\pi/2}^n$ symmetry.
Thus, in principle, 
only $n+1$ boundary conditions on equal classes of fixed points 
can be allowed to be different, if this symmetry is to be unbroken. 
Moreover, we can work out our analysis considering 
only the contribution of the
vacuum that resides on  the hypercubic slice ${\cal I}_0^n$.  Total potential
energy on the orbifold shall be just a $2^{n-1}$ multiple of this.

The solution, $\phi$, to the equation of motion (\ref{ec:mov}) on the flat
$n$-dimensional space we are considering 
can be factored as $\phi(\vec y)= \Pi_i^n\varphi_i(y_i)$,
where each independent factor  
is a solution to the generic equation $\varphi_i'' - k^2\varphi_i=0$, where 
$nk^2=\kappa^2=m^2b^2$, with the
boundary conditions $\varphi_i(0)=v_{i0}$ and $\varphi_i(1)= v_{i1}$, 
such that the  whole field configuration has boundary conditions
given by products of $v_i$'s. 
However, these $2n$ boundary conditions are not all independent. 
The ${\cal R}_{\pi/2}^n$ symmetry indicates that $v_{i1}v_{j0}$ is a 
constant for all  $i\neq j$, and so,
both $v_{i1}$ and $v_{i0}$ are independent of the index. 
This way, only two independent boundary conditions are actually needed, 
that we now choose as $v_{0,1}$, and thus,
all $\varphi_i$ would be the same  function  already 
given in Eq.~(\ref{general}), but evaluated for
the corresponding $y_i$ coordinate: 
$\varphi_i(y_i)=\varphi(y_i)=A e^{ky_i} + B e^{-ky_i}$, with the global 
constants 
$A=(v_1-v_0 e^k)/\sinh\,k$ and $B=(v_0e^{-k}-v_1)/\sinh\,k$. 
In this scenario different
directions along any coordinate axis look alike for the scalar field. That is 
the reason why volume varies as a whole while the basic geometry stands still.
We also note that $v_{0,1}^n$ should now
have mass dimension $1+n/2$ as the bulk scalar
field $\phi$.

The potential energy from this vacuum, as calculated in the initial frame is given now by the
general expression
$U_{ini}^n(b) = 2^{n-1}\times\frac{1}{2}b^{n-2}
\int_0^1\!dy\,\left[n(\varphi'(y))^2 +\kappa^2\varphi^2(y)\right]
\cdot \left[\int_0^1\!dy\,\varphi^2(y)\right]^{n-1}$. After some algebra, one
gets the rather complicated expression
 \bea
 U_{ini}^n(b) &=& \frac{n^{n/2}}{2m^{n-2}}
 \left(\frac{\left(v_0^2+v_1^2\right)\,\cosh\, k -2v_0v_1}{\sinh\, k}\right)
 \times\nonumber\\[1ex]&&
 \left(\frac{2v_0v_1\,\left(k\cosh\, k - \sinh\, k\right) + 
           \left(v_0^2+v_1^2\right)\,\left(\cosh\, k\, \sinh\,k -k\right)}
	   {\sinh^2\,k}\right)^{n-1}~,
 \eea
for which one can not stablish the existence for a minimum by exact analytical
methods. A numerical analysis, however,  shows that a minimum exist only for
$n=1$, which reduces to the case we discussed already in the previous section. One
can get some understanding for  
the reasons of this fact by looking at the behavior at small and
large $b$. For small radius one gets 
$U_{ini}^n\propto (1-a^2)^2(1+a+a^2)^{n-1}b^{n-2}$, 
where $a=v_0/v_1$, such that for $n=1$ it  diverges linearly as we already
know,  whereas it goes to a constant for $n=2$, and  to zero with a power
law for larger $n$. In contrast, for large radius 
the potential goes exponentially to a constant value $\propto (1+a^2)^n$.
Interpolating between these two extreme values with exponentially dominating
pieces, like those in the potential, 
leaves little room to develop any additional minimum.

As before, a minimum for the corresponding radion potential with $n>1$
may exist for some added configuration of bulk and brane cosmological constants. 
Consider once more the radion potential in 
Eq.~(\ref{urad}) with our present $U_{ini}^n$. It is clear
from the previous analysis that 
$U_{rad}^n\propto const./b^{2+n}+\lambda/b^{2n}$,
for small $b$, and thus, one would requires $\lambda>0$. On the other
hand, at large $b$ one gets $U_{rad}^n\propto \Lambda/b^n$. This is altogether
a similar behavior as the one already seen in the case of the interval for 
$v_0/v_1=1$ (the symmetric case). Nevertheless, 
here the conclusion arises regardless the
value of $v_0/v_1$ ratio. As before, a negative $\Lambda$ would be
enough to get a non trivial minimum, but at the unwanted
cost of a strictly negative value for $U_{rad}^n(b_0)$.

A more appealing scenario emerges if instead of $\lambda$ we assume that the
boundaries of the hypercube contribute to the potential energy with some surface
energy, fed by $(n-2)$-brane tensions. Thus,we add a potential term similar to the one
provided in Eq.~(\ref{tau}). Next we first consider 
the effective potential  written as
 \be
 U_{eff}^n = \left(\frac{b_0}{b}\right)
 \left[\left(\frac{b_0}{b}\right)^{n-1}U_{ini}^n(b) + \tau_n\right]~. 
 \ee  
The term between squared parenthesis in above equation still has no local 
minimum by itself, but now we can choose $\tau_n$ to insure that 
$U_{eff}^n$ will have one, 
by using a variation of the second strategy discussed by
the end of section three. First, notice that $U_{ini}^n/b^{n-1}$ goes as 
$\sim const./b$ for small $b$, so it is linearly divergent at zero.
Second, same term vanishes asymptotically as $\sim (v_0^2+v_1^2)^n/b^{n-1}$.
Hence, when shifting 
$U_{ini}^n/b^{n-1}$ by adding a negative $\tau$,  
we still get a function with no local minimum, which now,
however, crosses to negative values at some point, and approaches $\tau$ for
large $b$. Hence, the observation we made in previous section will apply: the
crossing is a fix point under the modulation by the overall $1/b$ factor, yet
to be included in order to build $U_{eff}^n$. 
As a matter of  fact, the multiplication by $1/b$ also 
changes the asymptotic form, and now 
$U_{eff}^n$ shall reach zero at infinity from below, as $\sim \tau_n/b$. 
Therefore, a local minimum, $b_0$, must now
emerge  within the region beyond the crossing point, where
$U_{eff}^n$ is negative. Finally, we shall 
consider a positive $\Lambda=-U_{eff}^n(b_0)$,
to shift the minimum value of 
 \be
 U_{rad}^n= \left(\frac{b_0}{b}\right)^n\left[\left(\frac{b_0}{b}\right)
 \left[\left(\frac{b_0}{b}\right)^{n-1}U_{ini}^n(b) + \tau_n\right] +
 \Lambda_n\right]~.
 \ee
to zero. A further  contribution of $\lambda$ is
not required now, although, it may be included too.

%%%%%%%%%%%%%%%%%%%%%
\begin{figure}%[ht]
%\vskip2em
\centerline{
\epsfxsize=220pt
\epsfbox{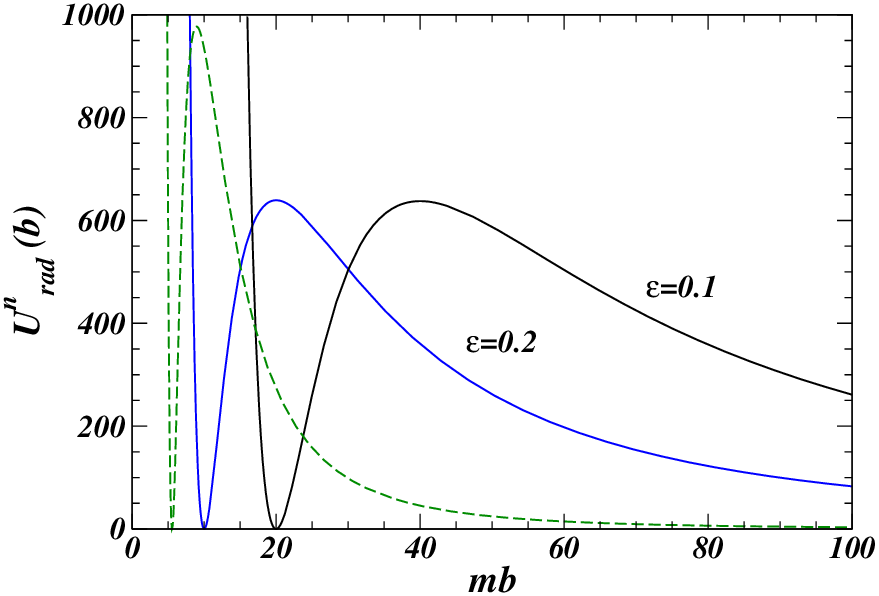}
}
%\vskip1ex
\caption{Radion stabilization 
potentials generated for the $T^n/Z_2$ orbifold, 
in units of $n^{n/2} v_1^{2n}/2m^{n-2}$ and for $v_0/v_1=10$. 
Continuous lines plot the profile  for $n=2$ and   
$\tau = -\varepsilon (1+v_0^2/v_1^2)^n$  
in same units as the potential, with $\varepsilon$ as indicated.
We also depict the profile for $n=3$ and $\varepsilon=0.1$ (dashed line)
with the  potential conveniently scaled by a factor of 1/100.}
\end{figure}
%%%%%%%%%%%%%%%%%%%

As an example we have chosen 
$\tau_n=-\varepsilon (1+v_0^2/v_1^2)^n\times(n^{n/2} v_1^{2n}/2m^{n-2})$
and made plots for the potential profile for $v_0/v_1 =10$ and for $n=2,3$;
using values of
$\varepsilon$ as  shown in figure~5. 
Notice again the characteristic form of these profiles, which interpose 
a potential barrier between the local minimum and infinity, and
whose width is actually sizable. 
Note also that the narrower  and taller 
barrier in the $n=3$ case (shown in the figure  
with an appropriate scaling factor to fit it within the used scale)
is actually an apparent effect due to the use of a numerically larger value of
$\tau$, although we are using the same value for $\varepsilon$.
This is also  the reason for which we now look at larger 
$mb_0$ values, when compared to previous figures.

%%%%%%%%%%%%%%%%%%%%%%%%%%%%%%%%%%%%%%%%%%%%%%%%%%%%%%%%%%%%%%%%%
\section{Conclusions}

Summarizing, our present work pin points a
clear conclusion: 
the combination of bulk and brane cosmological
constants and bulk vacuum energy from scalar fields does provide
successful and manageable scenarios for
the understanding of the stabilization of the radion field, within the context
of the four dimensional effective theory, in flat extra dimension models.
We have developed  
some basic strategies to handle 
and build radion potentials, with local minima and zero effective 
cosmological constant, out of the two above mentioned
minimal ingredients..   

Our analysis has been properly done in the Einstein frame, where
the radion is defined as a scalar field 
associated with volume variations and gravity is written in the standard form. 
We  properly included the volumetric suppressions 
introduced by conformal factors in all the different contributions to the 
radion potential we considered. 
We have shown that due to this factors, 
the use of a bulk cosmological constant and brane tension
configurations may be enough to provide  stabilization for the
radion. However, for the one extra dimension case,
an effective four dimensional negative cosmological constant arises. 

The further addition of a non trivial $y$-dependent vacuum energy
introduces the required freedom to obtain working scenarios for the
stabilization of the radion. These scenarios are good toy
models where other common problems of dynamical stabilization 
could be consistently analyzed, as other moduli stabilization or 
metric backreactions, that we have not discussed in here.
For example, a generalization of the present ideas to the stabilization of other
moduli fields is in principle possible. A trivial extension for the $T^n/Z_2$
orbifold may consider a separate stabilization of every each bulk direction,
using scalar fields located at the different boundaries on the orbifold, such
that the problem would get reduced to one dimensional cases. Other
configurations may also be possible. Backreactions, on the other hand, 
are less trivial to analyze  and it is a issue that still requires some study. 

Our results are an indication that it is well possible to built
phenomenological stabilization potentials, out of the most common ingredients
that any bulk-brane theory could have: brane and bulk cosmological constants,
and bulk scalar degrees of freedom with non trivial bulk configurations. 
We made no  claims on the possible size of the extra compact space,
but rather emphasize the fact that, 
even though, the size always appears related to the
scalar mass, in our constructions there are many possible situations
where the hierarchy on those parameters is conveniently sizable.
Nevertheless, such freedom usually means to move such hierarchy to the boundary
conditions on the scalar vacuum.

On the other hand, and mostly due to the conformal factors, 
all examples  we have provided suffer from the same potential illness: 
a decompactified extra dimensional volume appears also as a plausible scenario. 
We have not consider, however, any string correction nor quantum gravity effect
in our analysis. This has been so due to the very nature of our effective low
energy ($4D$) approach. We belive this problem might be ameliorated  
in a real  quantum gravity theory calculation, and it probably 
should not be a matter of concern in here. 
Moreover, close to the minimum and due to the Planck suppressions, our
model provides a workable scenario on which an effective theory approach should
properly describe radion physics. In particular, the approach may supply the
physical radion mass, characteristic of each particular model,  and certainly
the profile of the radion potential close to the minimum, too

As a final note, let us mention that because
cosmological constants contribute non trivially to the radion potential, 
any redefinition of those, either introduced by hand or due to quantum
contributions, may 
alter the stabilization of the volume in two possible ways. 
It may shift the minimum of the potential and introduce a non trivial
contribution to the effective 4D cosmological constant.
Intriguingly, this seems to stablish a connection of 
the cosmological constant hierarchy problem 
with the volume stabilization that may deserve further study.

%%%%%%%%%%%%%%%%%%%%%%%%%%%%%%%%%%%%%%%%%%%%%%%%%%%%%%%%%%%%%%%%%%

\section*{Acknowledgments}

ES and APL would like to thank the warm
hospitality of The Abdus Salam ICTP, at Trieste, Italy,
where part of this work was done. 
ES acknowledges CINVESTAV for
the hospitality and support along his many visits during the realization of this
work. 
Authors would like to thank to SIIN-UNACH06 by the partial support for this work.
APL and ES work was partially supported
by CONACyT, M\'exico, under grant J44596-F.

%%%%%%%%%%%%%%%%%%%%%%%%%%%%%%%%%%%%%%%%%%%%%%%%%%%%%%%%%%%%%%%%%%%%%

\end{document}